\documentclass[a4paper,11pt]{article}
\pdfoutput=1 
\usepackage{jheppub} 
\usepackage{epsfig}
\usepackage{caption}
\usepackage{yfonts}
\usepackage{subcaption}
\usepackage{color,soul}
\usepackage{slashed}
\usepackage{float}
\usepackage{amsfonts}
\usepackage{url}
\usepackage{slashed}
\usepackage{accents}
\usepackage{bbm,multicol}
\usepackage{lipsum}
\usepackage{mathtools}
\usepackage{physics}
\usepackage{enumitem}
\usepackage{array}
\usepackage{multirow}
\usepackage{booktabs}
\usepackage{makecell}

\newcommand{\crossLOzero}{{\langle \sigma v \rangle}_{\rm LO}^{m^{(0)}}}
\newcommand{\omegaLOzero}{\left( \Omega h^2 \right)^{m^{(0)}}_{\rm LO}}

\usepackage[normalem]{ulem}

\hypersetup{
  bookmarks=true,         % show bookmarks bar?
  unicode=false,          % non-Latin characters in Acrobat?s bookmarks
 pdftoolbar=true,        % show Acrobat?s toolbar?
 pdfmenubar=true,        % show Acrobat?s menu?
 pdffitwindow=true,     % window fit to page when opened
 pdfstartview={FitH},    % fits the width of the page to the window
 pdfsubject={Dark Matter},   % subject of the document
 pdfnewwindow=true,      % links in new window
 pdfcreator={RevTeX},
 colorlinks=true,       % false: boxed links; true: colored links
 linkcolor=red,          % color of internal links
 citecolor=blue,        % color of links to bibliography
 filecolor=black,      % color of file links
 urlcolor=blue,           % color of external links
  }

\title{Finite Temperature NLO Corrections in Relativistic Scatterings: Implications for Dark Matter Freeze-In}
\author[a]{Sampriti Roy,}
\author[a]{Pritam Sen,}
\author[a]{and Satyanarayan Mukhopadhyay}
\affiliation[a]{School of Physical Sciences, Indian Association for the Cultivation of Science (IACS), 2A and 2B Raja S.C. Mullick Road, Kolkata 700 032.}

\emailAdd{spssr2924@iacs.res.in}
\emailAdd{spsps3333@iacs.res.in}
\emailAdd{tpsnm@iacs.res.in}

\abstract{We study the next-to-leading order (NLO) virtual and thermal corrections to relativistic $2 \rightarrow 2$ scattering processes involving scalar particles in the early Universe thermal plasma. Taking the example of  freeze-in production of scalar dark matter pairs through these scatterings, we evaluate the impact of the NLO corrections to the annihilation rate and the dark matter yield. We find that including only thermal mass corrections to a leading order interaction rate can overestimate the reduction in these rates, and the full NLO corrections can modify the DM abundance predictions by $\mathcal{O}(30\%)$. It is also observed that while the virtual NLO effects are larger, the finite temperature NLO corrections to the matrix elements in the relativistic regime can modify the DM abundance by $\mathcal{O}(10\%)$, in comparison to the virtual NLO corrections.}

\begin{document} 
\maketitle
\flushbottom

\section{Introduction}
\label{sec:intro}
\raggedbottom

In studying particle production in the early Universe thermal plasma, the commonly adopted method is to compute the process matrix elements in vacuum quantum field theory (QFT) formalism, and subsequently average over the initial particle momentum distribution functions. This thermally averaged reaction rate is then incorporated in a Boltzmann kinetic equation to obtain the particle phase-space distribution functions, or the number densities. It is well-known that this method is an approximation to more accurate computations in a thermal field theory (TFT) framework, in which the particle production rates are encoded in correlation functions computed in TFT~\cite{Kapusta:2006pm, Bellac:2011kqa, Laine:2016hma}. In particular, important TFT corrections to the above commonly adopted method can come from several sources in a thermal plasma:
\begin{enumerate}[label=(\arabic*), nosep]
\item In a finite temperature ($T$) and density environment, the in-medium mass of the particles can differ significantly if they are sufficiently interacting with the medium. This effect is especially pronounced at high temperatures compared to the renormalized mass of the particle at $T=0$. The modified in-medium mass has an impact on the phase-space element in a scattering or decay process, as well as in the particle thermal distribution functions.

\item There can be additional stimulated emission or absorption processes of on-shell particles to or from the bath, contributing to the reaction rates.

\item Next-to-leading order (NLO) virtual corrections to the process matrix elements can become important in precision computations. At the same order in perturbation theory as these corrections, finite-$T$ NLO corrections to the matrix elements can also become relevant for accurate determinations of particle abundances in different scenarios.
\end{enumerate}
In the context of early Universe cosmology, there have been several  important studies in this direction. For particle decays in the early Universe, such as neutron or muon decay, which could be relevant in precision analyses of processes such as the Big Bang Nucleosynthesis, see, for example, Refs.~\cite{Dicus:1982bz, Cambier:1982pc, Baier:1989ub, Brown:2000cp, Czarnecki:2011mr}. Further important studies on thermal effects in neutrino decoupling were carried out in Refs.~\cite{Fornengo:1997wa, Bennett:2020zkv, Jackson:2023zkl}. For thermal matrix element corrections in particle scatterings, there have been studies in the context of thermal leptogenesis, see, for example, Refs.~\cite{Covi:1997dr, Giudice:2003jh, Anisimov:2010aq, Anisimov:2010dk, Beneke:2010wd, Salvio:2011sf, Besak:2012qm}, as well as in dark matter (DM) pair-production from thermal freeze-out (FO)~\cite{Wizansky:2006fm, Beneke:2014gla, Beneke:2016ghp, Butola:2024oia, Butola:2025vgg}. It was observed by Beneke et al.~\cite{Beneke:2014gla, Beneke:2016ghp} and Butola et al.~\cite{Butola:2024oia, Butola:2025vgg} that in the non-relativistic freeze-out of a fermion dark matter to Standard Model (SM) fermion pairs, through a $t$-channel scalar mediator, the NLO thermal corrections are suppressed at least by $\left(\frac{T_{F.O.}}{m_\chi}\right)^2$, where, $T_{F.O.}$ is the freeze-out temperature and $m_\chi$ is the DM mass. With the commonly observed value of $T_{F.O.}\sim m_\chi/20$, this amounts to a correction of around $0.25\%$. For several other important discussions on thermal effects in dark matter cosmology, we refer the reader to Refs.~\cite{Beneke:2016ync, Kim:2016kxt, Biondini:2017ufr, Biondini:2018xor, Biondini:2018pwp, Binder:2018znk, Biondini:2023zcz, Becker:2023vwd}, and references therein. 

Our study concerns primarily with the role of finite-$T$ NLO corrections, and its comparison with the vacuum NLO corrections in scattering processes involving relativistic particles. In particular, we analyze these corrections in the context of freeze-in production of dark matter (DM) in $2 \to 2$ scattering processes, in which the particles involved are relativistic, with $m^{(0)} < T$, $m^{(0)}$ being the $T=0$ renormalized mass parameters. Therefore, this study can be considered complimentary to the kinematic regime considered in Refs.~\cite{Beneke:2014gla, Beneke:2016ghp, Butola:2024oia, Butola:2025vgg}. Our primary objective is to determine whether the thermal NLO corrections for DM production in relativistic scatterings could be larger than those obtained for non-relativistic DM annihilations. 

To this end, we consider a simple toy model for dark matter ($\chi$) production through the scattering of a pair of relativistic SM-like particles ($\phi$) in the early Universe: $\phi\phi \to \chi\chi$. This process belongs to the category of so-called relativistic freeze-in processes~\cite{Hall:2009bx}. We study both the vacuum and thermal NLO corrections to the matrix elements of this process, and find out its impact on the DM yield. These corrections are not only relevant for precise computations of DM abundance within this model, they also serve as a prototype to how large the NLO thermal corrections could be in the relativistic regime.

The paper is organized as follows. In Sec.~\ref{sec:sec2} we describe the basic setup for computing the DM yield in a relativistic freeze-in scenario, while in Sec.~\ref{sec:sec3} we describe the computational details of the NLO corrections to the process $\phi\phi \to \chi\chi$, using the real-time framework for TFT. In Sec.~\ref{sec:sec4} we present numerical results regarding the NLO corrections and its impact on the DM yield. We summarize our findings in Sec.~\ref{sec:sec5}.

\section{Scalar dark matter freeze-in from scattering}
\label{sec:sec2}
We study in detail a simple DM scenario, with a real scalar DM ($\chi$), produced from the freeze-in of a pair of real scalar particles ($\phi$), where the latter are taken to represent SM-like fields, with a small mixing with the Higgs boson such that kinetic equilibrium with the SM bath can be maintained. We assume that at the end of post-inflationary reheating, the $\phi$ particles are populated, while the $\chi$ particles are not produced at this epoch due to negligible coupling to the inflaton field. The bare interaction Lagrangian is given by
\begin{equation}
\mathcal{L}_{\rm int} \supset -\frac{g}{4} \chi^2 \phi^2 -\frac{\lambda_{\chi}}{4!} \chi^4 - \frac{\lambda_{\phi}}{4!} \phi^4,
\label{Eq:int_lag}
\end{equation}
where all the parameters are real, and for simplicity only the above quartic couplings are considered. For a representative selection of earlier studies on freeze-in production of DM, see, for example, Refs.~\cite{Hall:2009bx, Drewes:2015eoa, Darme:2019wpd, Biondini:2020ric, Konig:2016dzg, Lebedev:2019ton, Arias:2020qty, DeRomeri:2020wng, Li:2023ewv, Koivunen:2024vhr,Becker:2023vwd}, and references therein.

In order to obtain relativistic freeze-in, consider a scenario in which the zero-temperature masses of the particles satisfy the hierarchy $m_\phi^{(0)}< m_\chi^{(0)}$, while due to a larger self-coupling, at high temperatures, the thermal masses satisfy $m_\phi^{(T)}>m_\chi^{(T)}$. We consider the DM sector to be non-thermal, and  populated only through freeze-in from thermal $\phi$ particle scatterings. With the $\phi-\chi$ coupling $g$ taking very small values, the DM particles never reach kinetic or chemical equilibrium with the $\phi$ particles in the bath. We take the $\phi$ particles to be in equilibrium with the SM sector with a temperature $T=T_{\rm SM}$.  Given the small DM couplings, the in-medium mass corrections of  $\chi$ is very small, such that for all practical purposes, $m_\chi \sim m_\chi^{(0)}$. We note that while $m_\phi^{(0)}< m_\chi^{(0)}$ and $m_\phi^{(T)}>m_\chi^{(T)}$ clearly facilitates relativistic freeze-in, this is not strictly necessary, as examples in the later sections will demonstrate.

Given the above setup, the dominant production of DM is expected to take place through the freeze-in process $\phi \phi \rightarrow \chi \chi$ at $T>T^\prime$, where $m_\phi^{(T)} > m_\chi^{(0)}$ is satisfied. Once $T<T^\prime$, the DM production becomes sub-dominant, and takes place only through the high-momentum tail of the $\phi$ distribution function. With $f_\chi<<1$ we can ignore back-reactions and Bose enhancement in DM production, in which case the DM number-density may be obtained by integrating the following approximate rate equation:
\begin{equation}
\frac{dn_\chi}{dt}+3H n_\chi \simeq {\langle \sigma v \rangle}_{\phi \phi \rightarrow \chi \chi} n_\phi^2,
\end{equation}
where, we take $n_\phi$ to have a thermal distribution. With the standard redefinitions, $Y = n/s$, and $x=m_\chi^{(0)}/T$, with $s$ being the entropy density of the Universe, we have the following evolution equation for $Y_\chi$:
\begin{equation}
\frac{dY_\chi}{dx}\simeq \frac{s {\langle \sigma v \rangle}_{\phi \phi \rightarrow \chi \chi}}{xH} Y_\phi^2.
\end{equation}
Given the functional form of ${\langle \sigma v \rangle}_{\phi \phi \rightarrow \chi \chi}(x)$, we can integrate the above equation to obtain the number density of DM particles produced through freeze-in. The integration needs to be performed starting from  $x_{\rm min}=m_\chi^{(0)}/T_{\rm max}$, where, $T_{\rm max}$ is the maximum temperature at which we have a thermal density of $\phi$ particles available for annihilation. This is essentially determined by the reheating temperature, assuming that $\phi$ is in equilibrium with the SM bath across all relevant temperatures. For our numerical analysis, we have taken $x_{\rm min} \sim 1/100$. 

Satisfying the observed relic density through this freeze-in process leads to a required value of the $\phi$ annihilation rate, which is found to be very small. We also check whether this value of the annihilation rate is consistent with the out-of-chemical-equilibrium assumption for the DM throughout its thermal history, by approximately requiring $n_\phi {\langle \sigma v \rangle}_{\phi \phi \rightarrow \chi \chi} < H$. 

\section{DM freeze-in from scattering: virtual and thermal NLO corrections}
\label{sec:sec3}
\begin{figure}[H]
\centering
\includegraphics[width=0.20\textwidth]{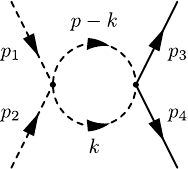}~~~~~~~~~
\includegraphics[width=0.20\textwidth]{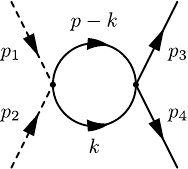}  
\includegraphics[width=0.20\textwidth]{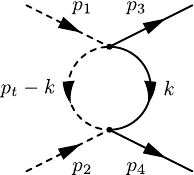} 
\includegraphics[width=0.20\textwidth]{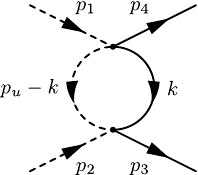}~~~~~~~~~ 
\caption{\em{Feynman diagrams representing the virtual and thermal  NLO corrections to the matrix elements for the freeze-in process $\phi \phi \rightarrow \chi \chi$, with the momentum labels $p=p_1+p_2, ~p_t=p_1-p_3$ and $p_u=p_1-p_4$. The dashed and solid lines represent $\phi$ and $\chi$, respectively.}} 
\label{Fig:fig1}
\end{figure}

Having discussed the basic setup for scalar DM relativistic freeze-in in the previous section, we now move onto the analysis of higher order NLO corrections to the scattering rate ${\langle \sigma v \rangle}_{\phi \phi \rightarrow \chi \chi}(x)$, both from virtual NLO effects, as well as from thermal NLO corrections. In Fig.~\ref{Fig:fig1} we show the Feynman diagrams at next-to-leading order for the process $\phi \phi \rightarrow \chi \chi $. As noted in the previous section although $g << {O}(1)$ is expected from the point-of-view of satisfying the DM relic abundance through freeze-in, however, the self-couplings $\lambda_\phi$ and $\lambda_\chi$ are restricted only by perturbativity at this stage, and therefore can be sizeable. We therefore impose only the weak upper bound of $\lambda_{\phi,\chi} \leq \sqrt{4\pi}$ in this study. Furthermore, since we are considering the regime $g << \{\lambda_\phi, \lambda_\chi \}$, the $t$- and $u$- channel diagrams in Fig.~\ref{Fig:fig1} are highly suppressed compared to the two $s$-channel diagrams. Therefore, in this study, we shall be focussing on the NLO contributions from the $s$-channel diagrams involving either the $\phi$ particle loop or the $\chi$ particle loop only.

We compute the NLO thermal corrections to the matrix elements (ME) of the $\phi \phi \rightarrow \chi \chi$ process using the real-time formalism of thermal field theory (TFT)~\cite{Bellac:2011kqa}. In the simple one-loop example being considered here, with all the vertices connected to external physical bath particles, the only relevant component of the propagator is as follows~\cite{Bellac:2011kqa, Czarnecki:2011mr, Nishikawa:2003js}: 
\begin{equation}
D_{11}(k) = \frac{i}{k^2-m(T)^2+i\epsilon}+ 2 \pi f(k_0) \delta(k^2-m(T)^2),
\label{Eq:prop}
\end{equation}
where, we have used 1PI resummed propagator~\cite{Giudice:2003jh}, and $m(T)^2$ is the effective plasma mass squared of the scalar particle~\footnote{Instead of the 1PI resummed propagator one could have also used a daisy resummed propagator~\cite{Kapusta:2006pm, Bellac:2011kqa, Laine:2016hma, Dolan:1973qd}. We have explicitly checked that the resulting shift in the thermal mass value for our choice of parameters is at most $\mathcal{O}(3\%)$ of the 1PI thermal mass in the relativistic regime. Therefore, in what follows, we have neglected the daisy resummation effect in our numerical analysis.}, which includes the bare mass squared, and the Bose-Einstein distribution function $f(k_0)$ is given by
\begin{equation}
f(k_0) = \frac{1}{e^{\beta |k_0|}-1},
\end{equation}
with $\beta=1/T$ being the inverse bath temperature. In writing Eq.~\ref{Eq:prop}, we have neglected the absorptive part of the scalar boson resummed propagators\footnote{The absorptive part only arises at two-loop order in this theory from scalar sunset type diagrams, and is thus neglected while calculating results till this order~\cite{Nishikawa:2003js}.}, as they are suppressed compared to thermal plasma mass correction~\cite{Giudice:2003jh}.  

Although generically in the real time formalism, both type 1 and type 2 vertices exist~\cite{Bellac:2011kqa}, vertices linked to external momenta can only be of type 1, while internal vertices can be of either type. As mentioned above, at the one-loop order diagrams with scalar quartic interactions considered in this study (see Fig.~\ref{Fig:fig1}), the only possible vertices are of type 1, and therefore, only the $D_{11}$ propagator connecting these appear, the details of which is outlined in the Appendix. In the two-loop diagrams for 2 to 2 scattering processes, however, all the four components of the propagator become relevant.

In addition to the NLO thermal corrections to the ME's of $\phi \phi \rightarrow \chi \chi$, this modified propagator impacts the evolution of DM number density in a thermal bath in several ways. Self-energy diagrams now lead to both quantum and thermal corrections to the particle masses, where the thermal mass correction component for $\phi$ is given by:
\begin{equation}
m_\phi^2(T) - ({m^{(0)}_\phi})^2 \simeq \frac{\lambda_\phi}{2} \int \frac{d^3 k}{(2\pi)^3} \frac{1}{E_{\vec{k}}}  f_\phi (E_{\vec{k}}),
\label{Eq:thermal_mass}
\end{equation}
where, ${m^{(0)}_\phi}$ represents the zero-temperature renormalized mass of the $\phi$ particle. While we use the general mass correction formula in Eq.~\ref{Eq:thermal_mass} in our numerical analysis, for analytical approximations, a particular simple expansion can be made for example in the limit ${m^{(0)}_\phi}<<T$, with,
\begin{equation}
m_\phi^2(T) - ({m^{(0)}_\phi})^2 \equiv \delta m^2_{T,\phi} \simeq \frac{\lambda_\phi}{2} \left(\frac{T^2}{12}-\frac{m_\phi^{(0)} T}{4\pi}\right).
\end{equation}
This modification in the particles masses will modify the phase-space elements, as well as the distribution functions which depend on the mass. In particular, for example, for a fixed temperature, the $\phi$ number density will now be reduced. We shall discuss the impact of these thermal corrections systematically in a subsequent section. 

We now discuss the $\phi \phi \rightarrow \chi \chi$ amplitude in detail, taking the s-channel diagram in Fig.~\ref{Fig:fig1} with $\phi$ particles in the loop as an example. This Feynman diagram represents the NLO amplitude
\begin{equation}
i \mathcal{M^{\phi \phi}_{\rm NLO}} = - \frac{g \lambda_\phi}{2} \left( M_{\rm VV} + M_{\rm VT} + M_{\rm TT} \right),
\end{equation}
where the terms represent the loop integrals $M_{\rm IJ} = \int \frac{d^4 k}{(2\pi)^4} \, \widetilde{M}_{\rm IJ}$, with,
\begin{equation}
\begin{aligned}
\widetilde{M}_{\rm VV} &= \frac{-1}{(k^2 - m^2 + i \epsilon) \left( (p - k)^2 - m^2 + i \epsilon \right)}, \\
\widetilde{M}_{\rm VT} &= \frac{i}{k^2 - m^2 + i \epsilon} \, 2 \pi f(p^0 - k^0) \, \delta((p - k)^2 - m^2)+\frac{i}{(p - k)^2 - m^2 + i \epsilon} \, 2 \pi f(k^0) \, \delta(k^2 - m^2), \\
\widetilde{M}_{\rm TT} &= 4\pi^2 \, f(k^0) \, f(p^0 - k^0) \, \delta(k^2 - m^2) \, \delta((p - k)^2 - m^2),
\end{aligned}
\label{Eq:NLO_amp}
\end{equation}
where, we have written $m^2=m_\phi^2(T)$ for brevity. Here, the last term in Eq.~\ref{Eq:NLO_amp}, $M_{\rm TT}$ corresponds to the contribution of two on-shell bath particles in the loop diagram, which has already been included at this order in perturbation theory by the leading order ME of $\phi \phi \rightarrow \chi \chi$, coupled with elastic scatterings. Therefore, at this coupling order, we drop the  ${M}_{\rm TT}$ term, to avoid double counting of this contribution. $M_{\rm VV}$ represents the vacuum NLO correction from the $s$-channel diagrams. For example, with the $\phi$ particle in the loop, this contribution is given by~\cite{Peskin:1995ev}:
\begin{equation}
M_{\rm VV}
\supset
-\frac{i}{16\pi^2}
\sqrt{\frac{s-4m^2}{s}}
\;
\log
\!\left(
\frac{2m^2-s+\sqrt{s(s-4m^2)}}
     {2m^2}
\right),
\end{equation}
where we have renormalized the theory in the on-shell scheme at $T=0$, at the renormalization point $s=4 {(m_\phi^{(0)}})^2$, $t=u=0$, where $s,t,u$ represent the standard Mandelstam variables. Similarly, there will be a corresponding s-channel vacuum NLO contribution from the $\chi$ particle loop as well. Since there are no additional sources of ultraviolet divergences at finite temperature, the UV renormalization of the theory remains unchanged.

The thermal contribution at NLO comes from the second term ${M}_{\rm VT}$. In order to obtain a closed form analytical expression for this term, it is useful to first separate the principal part of the propagator factors~\cite{Nishikawa:2003js}:
\begin{equation}
 \frac{i}{(k^2 - m^2 + i \epsilon)} = P\left(\frac{i}{k^2 - m^2}\right)+\pi \delta(k^2 - m^2),
\end{equation}
thus splitting the two contributions ${M}_{\rm VT}={M}_{\rm VT}^{(1)} + {M}_{\rm VT}^{(2)}$. The delta function piece can now be combined with the thermal delta function to yield the integrated closed form result:
\begin{equation}
{M}_{\rm VT}^{(2)} = \frac{1}{16\pi |\vec{p}| \beta} \ln \left|\frac{1-e^{-\beta \omega^+}}{1-e^{-\beta \omega^-}}\right|, 
\end{equation}
where, the limits on the energy integration are restricted in the range given by
\begin{equation}
\omega_{\pm} = \frac{p_0}{2} \left(1 \pm \frac{ |\vec{p}|}{p^0} \sqrt{1-\frac{4m^2}{p^2}}\right).
\end{equation}
Here, the total initial energy and three momentum in the s-channel have been defined as $p^0=p_1^0+p_2^0$, and $\vec{p}=\vec{p_1}+\vec{p_2}$, respectively. The other loop integral given by
\begin{equation}
{M}_{\rm VT}^{(1)} = \int \frac{d^4 k}{(2\pi)^4} P\left(\frac{i}{(p-k)^2 - m^2}\right) 2 \pi f(k^0) \, \delta(k^2 - m^2)
\label{Eq:thermal_piece1}
\end{equation}
cannot be computed exactly in closed form to our knowledge. Therefore, we perform the integral in Eq.~\ref{Eq:thermal_piece1} numerically in all our subsequent results. For certain kinematic regimes, an approximate analytical computation may be done by first expanding the integrand in a suitably defined power series, and then restricting the integral to that kinematic domain, as detailed in the next subsection. Such a method provides the approximate analytical form of the thermal NLO corrections in that domain, which helps in understanding how large the corrections could be.

\subsection{Analytical approximation for the thermal NLO matrix element}
In order to obtain an analytic closed form approximation of Eq.~\ref{Eq:thermal_piece1}, we note that since $f(E_{\vec{k}})$ is very small for $E_{\vec{k}}>T$, the dominant contribution to the integral over $\vec{k}$ comes from the region $|\vec{k}|\leq T$, where $E_{\vec{k}}=\sqrt{|\vec{k}|^2+m^2}$. After performing the $k_0$ integral in Eq.~\ref{Eq:thermal_piece1}, it is easy to see that we might expand the integrand as a power series as long as 
\begin{equation}
\left|\frac{2p.k}{p^2}\right| < 1.
\end{equation}
This condition needs to be satisfied for both $k_0=\pm E_{\vec{k}}$, which are the roots of the delta function argument.

Consider the case with $k^0=E_{\vec{k}}>0$. From the boundedness of $\cos \theta$ in $\vec{p}.\vec{k}=|\vec{p}||\vec{k}|\cos \theta$, and the fact that $ 4m^2 \leq p^2 < \infty$, we have the inequality 
\begin{equation}
\frac{2p.k}{p^2} < \frac{p^0 E_{\vec{k}}+(|\vec{p_1}|+|\vec{p_2}|)|\vec{k}|} {2m^2}.
\end{equation}
We have further used the triangle inequality to write $|\vec{p}| \leq |\vec{p_1}|+|\vec{p_2}|$ above. Now, as argued above, the integral receives its dominant contribution in the region $E_{\vec{k}}<T$, and consequently, $|\vec{k}|=\sqrt{E_{\vec{k}}^2-m^2}<T$. In this region, we then have
\begin{equation}
\frac{2p.k}{p^2} < \frac{{p_1^0+p_2^0}+|\vec{p_1}|+|\vec{p_2}|} {2m^2} T.
\end{equation}
The magnitude of this upper bound can now be evaluated for different possible initial state configurations, as well as the loop particle involved.

When the initial particles are non-relativistic, with $|\vec{p_i}|<<m_i$, $p_i^0 \simeq m_i$, the upper bound is approximately $T/m$. Now assume that the initial particles are $\phi$. In that case, non-relativistic $\phi$ also implies $T<m$. If the loop particle involved is $\phi$ as well, this leads to a contradiction, since the $E_{\vec{k}}<T$ condition can now no longer be satisfied. However, if the loop particle involved is a $\chi$, all the conditions can be satisfied, and we can have $\left|\frac{2p.k}{p^2}\right| < 1$. While this kinematic region is interesting from the point of view of non-relativistic DM freeze-out, it is not the most relevant one for the relativistic freeze-in scenario being studied here.

On the other hand, if the initial particles are relativistic, we have $|\vec{p_i}| \geq m_i$ and we can approximate  $p_i^0 \simeq |\vec{p_i}|+\frac{m^2}{2|\vec{p_i}|}$. Since the average value of  $|\vec{p_i}| \sim 3 T$ in the bath for a relativistic particle, the upper bound on $\left|\frac{2p.k}{p^2}\right|$ is then approximately $6T^2/m^2+1/6$. Therefore, as long as $|\vec{p_i}| \sim m_i \sim 3T$, this translates to an upper bound of $5/6$. Thus, we can expand the integrand in $\left|\frac{2p.k}{p^2}\right|$ in the relativistic region with $T \sim m_i/3$, but not for any higher temperature. For $k^0=-E_{\vec{k}}<0$, a similar argument shows that this upper bound is now stronger, namely, $1/6$, thereby ensuring a faster convergence of the Taylor expansion.  

With the above results, we can now expand the integrand in Eq.~\ref{Eq:thermal_piece1} as a power series in $\frac{2p.k}{p^2}$, and evaluate it analytically term by term. For example, the leading term in the series is obtained to be:
\begin{equation}
{M}_{\rm VT}^{(1)} \simeq \frac{i}{p^2} F(T,m),
\label{Eq:series_s}
\end{equation}
where, $F(T,m)$ is the well-known thermal integral:
\begin{equation}
F(T,m) = \int \frac{d^3 k}{(2\pi)^3} \frac{1}{E_{\vec{k}}} f(E_{\vec{k}}).
\label{Eq:F_th}
\end{equation}
Depending upon the mass of the particle appearing in $F(T,m)$, an analytical non-relativistic or relativistic expansion of this function may also often be performed. Thus, to summarize, in our freeze-in example, we can approximate the NLO thermal matrix element correction in the region $T \sim m_\phi/3$ by Eq.~\ref{Eq:series_s}, which implies an approximate scaling of this term as $T^2/s$ (where, $s=p^2$). This is consistent with the non-relativistic limit discussed in Ref.~\cite{Beneke:2014gla, Beneke:2016ghp, Butola:2024oia, Butola:2025vgg}, where the NLO thermal corrections were shown to scale as $T^2/m^2$, unless this term is suppressed due to selection rules. We have checked by comparison with exact numerical results that Eq.~\ref{Eq:series_s} gives a reasonable approximation in the region $T \sim m_\phi/3$ and somewhat, but not too much, smaller. However, these analytical approximations are discussed here only for a qualitative understanding of the thermal NLO corrections. In the subsequent section, we have used exact numerical integrals for the matrix elements without any approximations.

\section{Numerical results: NLO corrections to relativistic freeze-in}
\label{sec:sec4}
\begin{figure}[htb!]
    \centering
    \includegraphics[width=0.475\textwidth]{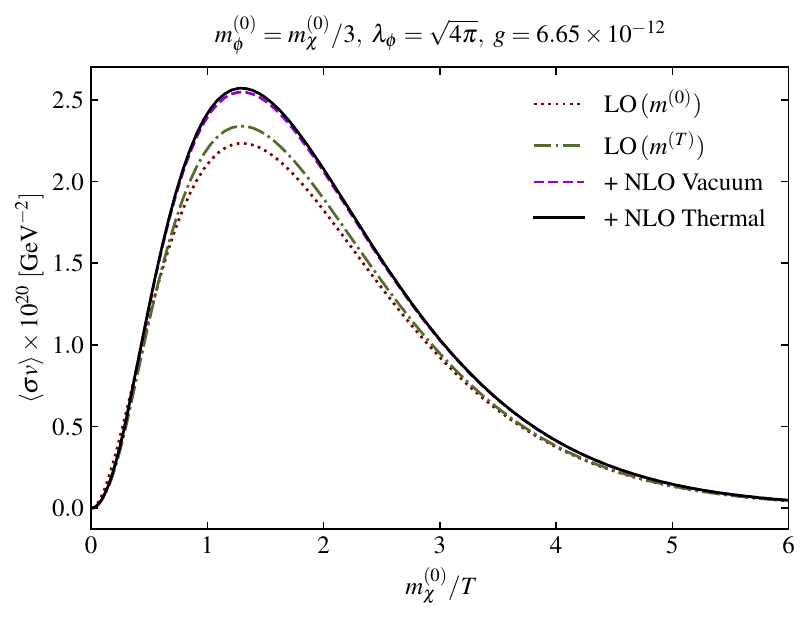}%
    \hspace{0.04\textwidth}%
    \includegraphics[width=0.485\textwidth]{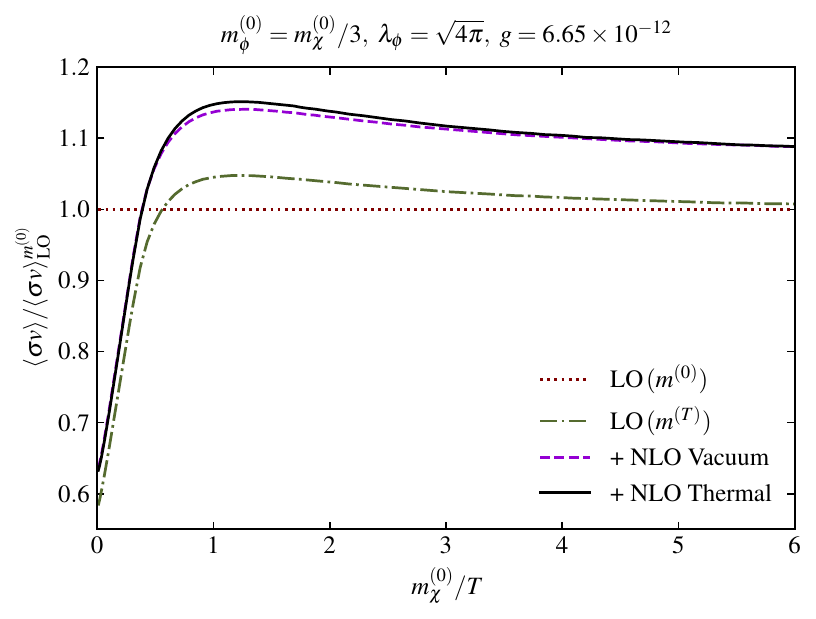}
    \caption{\em{(Left) The thermally averaged annihilation rate 
${\langle \sigma v \rangle}_{\phi \phi \rightarrow \chi \chi}(x)$ for the freeze-in process as a function of $x=m_\chi^{(0)}/T$,
(i) at LO, 
(ii) at LO including thermal mass effects, 
(iii) at NLO including only the vacuum NLO matrix elements and thermal mass, and 
(iv) at NLO including both the vacuum and thermal NLO matrix elements and mass corrections. 
(Right) The ratio of different annihilation rates with 
${\langle \sigma v \rangle}_{\rm LO}^{m^{(0)}}$.}}
    \label{fig:FI_1}
\end{figure}

We now present the results of numerical integration for the reaction rates at LO and NLO order, as well as the respective DM yields. In Fig.~\ref{fig:FI_1} (left panel), we show the thermally averaged annihilation rate 
${\langle \sigma v \rangle}_{\phi \phi \rightarrow \chi \chi}(x)$ for the freeze-in process as a function of $x=m_\chi^{(0)}/T$ for four different cases: (i) at LO, ${\langle \sigma v \rangle}_{\rm LO}^{m^{(0)}}$, (ii) at LO, but including thermal mass effects, ${\langle \sigma v \rangle}_{\rm LO}^{m^{(T)}}$, (iii) at NLO including only vacuum NLO matrix elements, with thermal mass effects, ${\langle \sigma v \rangle}_{\rm NLO}^{\rm V} $ and (iv) at NLO including both the vacuum and thermal NLO matrix elements and mass corrections, ${\langle \sigma v \rangle}_{\rm NLO}^{\rm T}$. We show all four of these cases in order to illustrate the relative impact of the four effects on the annihilation rate. While the easiest one to compute is ${\langle \sigma v \rangle}_{\rm LO}^{m^{(0)}}$, often in many studies that is improved by including the thermal mass effect to obtain ${\langle \sigma v \rangle}_{\rm LO}^{m^{(T)}}$, especially if kinematic thresholds become relevant. The consistent way to incorporate the thermal effects is of course to compute the reaction rate at NLO or higher order in perturbation theory, including the vacuum and thermal matrix element corrections, which gives ${\langle \sigma v \rangle}_{\rm NLO}^{\rm T}$. In order to determine the relative role of the vacuum and thermal NLO matrix element corrections, we also compute 
${\langle \sigma v \rangle}_{\rm NLO}^{\rm V} $.  In the DM literature, there exist several studies on either vacuum NLO corrections, or thermal NLO corrections separately, but the number of studies combining both the NLO effects together are fewer, although they occur at the same order in perturbation theory. On the right panel of Fig.~\ref{fig:FI_1}, we show the ratio of different annihilation rates with ${\langle \sigma v \rangle}_{\rm LO}^{m^{(0)}}$.

As we can see from these figures, all the reaction rates first increase upto a maximum around $x \sim 1.5$, before falling down. We also observe from the right panel figure that the other three rates are smaller than $\crossLOzero$ in a small range with $x<1$, but are all larger than $\crossLOzero$ for higher $x$ values. The difference between $\crossLOzero$ and ${\langle \sigma v \rangle}_{\rm LO}^{m^{(T)}}$ vanishes at larger $x$, as the thermal mass corrections become negligible there. However, a significant difference between $\crossLOzero$ and ${\langle \sigma v \rangle}_{\rm NLO}^{\rm V} $ or ${\langle \sigma v \rangle}_{\rm NLO}^{\rm T}$ remains for all $x$, due to the vacuum NLO effects. 

The above discussed features of Fig.~\ref{fig:FI_1} may be understood as follows. There are two competing effects at play here: (i) the $1/E_1 E_2$ factor from the flux entering the cross-section, which decreases the rate at high $T$ as the average particle energies become higher, and (ii) the phase-space factor in the cross-section, which is enhanced at higher $T$ due to the larger thermal corrections to $\phi$ mass, $m_\phi (T) > m_\chi(T)$. At very high $T$, the competition is won by the flux suppression, while for intermediate $T$ it is won by the phase-space enhancement, thereby explaining the features. For much lower $T$, of course the finite T corrections become negligible, and only the vacuum NLO corrections remain sizeable.

\begin{figure}[htb!]
    \centering
    \includegraphics[width=0.49\textwidth]{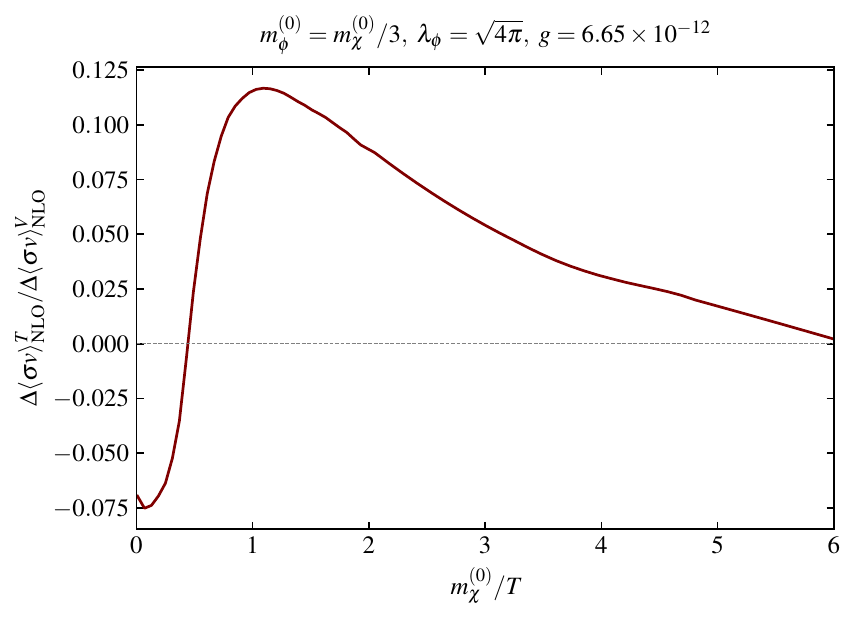}%
    \hspace{0.04\textwidth}%
    \includegraphics[width=0.47\textwidth]{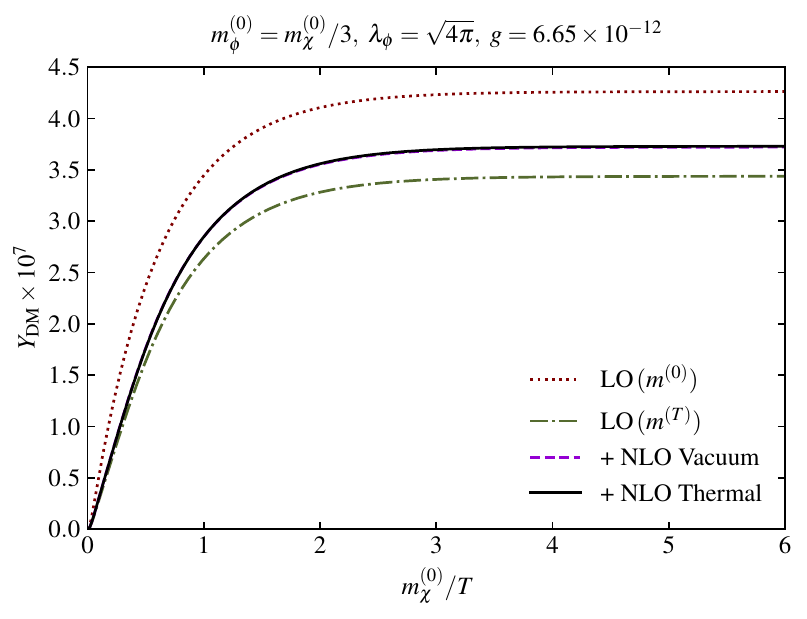}
    \caption{\em{(Left) Ratio of the reaction rates $\Delta \langle \sigma v \rangle_{\rm NLO}^{\rm T}/{\Delta \langle \sigma v \rangle_{\rm NLO}^{\rm V}}$ as defined in Eq.~\ref{Eq:ratio}, as a function of $x=m_\chi^{(0)}/T$. This ratio is representative of the thermal NLO corrections to the reaction matrix elements, see text for details. (Right) Dark matter yield $Y_{\rm DM}=n_\chi/s$, as a function of $x$, in all the four different computational setups considered.}}
    \label{fig:FI_2}
\end{figure}

In order to understand the impact of the thermal NLO correction matrix elements, it is instructive to study the following ratio:
\begin{equation}
\frac{\Delta \langle \sigma v \rangle_{\rm NLO}^{\rm T}}{\Delta \langle \sigma v \rangle_{\rm NLO}^{\rm V}} \equiv \frac{{\langle \sigma v \rangle}_{\rm NLO}^{\rm T} - {\langle \sigma v \rangle}_{\rm NLO}^{\rm V} } {{\langle \sigma v \rangle}_{\rm NLO}^{\rm V} - {\langle \sigma v \rangle}_{\rm LO}^{m^{(T)}}}.
\label{Eq:ratio}
\end{equation}
This ratio captures the effect of only the thermal NLO matrix element corrections, as all of the reaction rates involved are computed here including thermal masses, and therefore the common plasma mass effects are essentially subtracted out. We show in Fig.~\ref{fig:FI_2} (left panel) the ratio $\Delta \langle \sigma v \rangle_{\rm NLO}^{\rm T}/{\Delta \langle \sigma v \rangle_{\rm NLO}^{\rm V}}$, as a function of $x=m_\chi^{(0)}/T$. As we can observe from this figure, the thermal NLO matrix element corrections can be substantial in this relativistic scattering scenario, ranging from around $-8\%$ at very high $T$ to $+12\%$ at intermediate $T$. 

Finally, we study how these scattering rate differences translate to the DM yield $Y$ (we recall that $Y=n_\chi/s$) through the freeze-in process. We show in Fig.~\ref{fig:FI_2} (right panel) the yield $Y$ as a function of $x$, for all the four different computational setups considered above. First of all, we confirm from this figure that the freeze-in is dominated by the relativistic regime, as expected in earlier discussions, with most of the production happening at $x \lesssim 2$. Secondly, it is found that only including the thermal mass effect to the LO reaction rate gives a substantial reduction in the DM yield, due to the larger suppression from the flux factor at higher $T$ from the thermal masses, as well as from the reduction in $\phi$ distribution functions at a given $T$. However, simply including the thermal mass in the LO rate equation substantially overestimates the actual yield reduction at NLO, as seen from Fig.~\ref{fig:FI_2}. In particular, including the vacuum and thermal NLO corrections gives a prediction for the yield which falls in between these two different LO estimates. Therefore, we conclude from this figure that for precision determinations of DM yield in these relativistic freeze-in scenarios from scatterings, it is necessary to include the vacuum and thermal NLO effects.

In Table~\ref{Tab:tab-1}, we show the relative (percentage) change in relic abundance with respect to the leading order prediction $\omegaLOzero$, on including (i) thermal mass effects, and (ii) NLO vacuum matrix elements, and (iii) NLO thermal matrix elements. We also show the impact of NLO thermal matrix element corrections by the ratio defined in the last column. The results are shown for different ratios of the DM $\chi$ and the scalar $\phi$ mass, for two different choices of the $\phi$ self-coupling $\lambda_\phi = \{\sqrt{4\pi},1\}$, with the value of the $\chi-\phi$ interaction coupling $g$ fixed by demanding $\omegaLOzero$ matches the observed DM relic abundance through freeze-in. While computing the relic abundance, a DM mass input is needed, which we have fixed at $m_\chi^{(0)}=1$ MeV. Increasing this mass will require a corresponding decrease in the value of the coupling $g$, to keep $\omegaLOzero$ fixed. Such a reduction in $g$, as we have discussed, will have no impact on the $s$-channel vacuum and thermal NLO corrections being studied here. We note in passing that even if the DM is produced relativistic, its momenta will subsequently redshift due to the expansion of the Universe. This redshift can eventually make it non-relativistic at the epoch of matter-radiation equality, as required by structure formation. A sufficient redshift can be ensured by having earlier production epochs, and therefore, larger masses in our scenario. 
\begin{table}[htb!]
\renewcommand{\arraystretch}{1.5}
\hspace{-48mm}
\begin{center}
\scalebox{0.815}{\begin{tabular}{|c|p{0.8cm}|c|c|c|c|c|c|}  
\hline 
\rule{0pt}{40pt} $m_\phi^{\left(0\right)}/m_\chi^{\left(0\right)}$ 
& $ \multirow{1}{=}{\centering $\lambda_\phi$} $ & $g$
& $ \makecell{\dfrac{ \left( \Omega h^2 \right)^ {\rm m^{(T)}}_{\rm LO}}{\left( \Omega h^2 \right)^{\rm m^{\left(0\right)}}_{\rm LO} } -1 \\ [18pt] (\%)} $
& $ \makecell{\dfrac{ \left( \Omega h^2 \right)^ {\rm V}_{\rm NLO}}{\left( \Omega h^2 \right)^{\rm m^{\left(0\right)}}_{\rm LO} } -1 \\ [18pt] (\%)}$
& $ \makecell{ \dfrac{ \left( \Omega h^2 \right)^ {\rm T}_{\rm NLO}}{\left( \Omega h^2 \right)^{\rm m^{\left(0\right)}}_{\rm LO}} -1 \\ [18pt] (\%)} $ & $  \makecell{ \dfrac{ \left( \Omega h^2 \right)^ {\rm T}_{\rm NLO} - \left( \Omega h^2 \right)^ {\rm V}_{\rm NLO} }{\left( \Omega h^2 \right)^ {\rm V}_{\rm NLO} - \left( \Omega h^2 \right)^{\rm m^{\left(T\right)}}_{\rm LO}}  \\ [18pt] (\%) } $  \\ [25pt]
\hline

$1/3$ & \multirow{4}{=}{\centering $\sqrt{4\pi}$}  & $6.65 \times 10^{-12}$ & $  -19.36  $ & $  -12.69  $ & $  -12.50  $ & $   ~~~2.76  $  \\

\cline{1-1} \cline{3-7}

$1$ & & $7.40 \times 10^{-12}$ & $  -18.53  $ & $  -13.16  $ & $  -13.50  $ & $  -6.27  $ \\

\cline{1-1} \cline{3-7}
$2$ & & $1.00 \times 10^{-11}$ & $  -18.39  $ & $  -13.50  $ & $  -13.91  $ & $  -8.39  $ \\

\cline{1-1} \cline{3-7}
$5$ & & $1.65 \times 10^{-11}$ & $  -16.38  $ & $  -11.62  $ & $  -12.05  $ & $  -9.00   $ \\

\hline
$1/3$ & \multirow{4}{=}{\centering $1$} & $6.65 \times 10^{-12}$ & $  -8.38   $ & $  -5.86   $ & $  -5.72   $ & $   ~~~5.83  $ \\

\cline{1-1} \cline{3-7}

$1$ & & $7.40 \times 10^{-12}$ & $  -7.75   $ & $  -5.84   $ & $  -5.95   $ & $  -6.22  $  \\

\cline{1-1} \cline{3-7}
$2$ & & $1.00 \times 10^{-11}$ & $  -7.58   $ & $  -5.85   $ & $  -6.00    $ & $  -9.02  $ \\

\cline{1-1} \cline{3-7}
$5$ & & $1.65 \times 10^{-11}$ & $  -6.37   $ & $  -4.73   $ & $  -4.89   $ & $  -9.91  $ \\

\hline
\end{tabular}}
\renewcommand{\arraystretch}{1} \newline
\end{center}
\caption{\em{Relative (percentage) change in DM relic abundance with respect to the leading order prediction $\omegaLOzero$, on including (i) thermal mass effects, and (ii) NLO vacuum matrix elements, and (iii) NLO thermal matrix elements. We also show the impact of NLO thermal matrix element corrections by the ratio defined in the last column. The results are shown for different ratios of the DM $\chi$ and the scalar $\phi$ mass, for $\phi$ self-coupling $\lambda_\phi = \{\sqrt{4\pi},1\}$, with the value of the $\chi-\phi$ interaction strength $g$ fixed by demanding $\omegaLOzero$ matches the observed DM relic abundance through freeze-in.}}
\label{Tab:tab-1}
\end{table}

As we can see from Table~\ref{Tab:tab-1}, increasing the ratio $m_\phi^{\left(0\right)}/m_\chi^{\left(0\right)}$ leads to a higher required value of $g$. This is because, the increase in $m_\phi^{\left(0\right)}$ leads to a higher suppression to the reaction rate from both the flux factor and the $\phi$ distribution function, thereby requiring a larger couplings to keep $\omegaLOzero$ fixed. The absolute percentage effect of including the thermal mass in the LO rate is also slightly reduced as $m_\phi^{\left(0\right)}/m_\chi^{\left(0\right)}$ increases. The reason for this effect is that, for a fixed $\chi$ mass, as we make $\phi$ heavier, the DM yield receives substantial contributions from correspondingly higher values of $x$, for which the thermal mass corrections are smaller. Therefore the integrated effect on $\Omega h^2$ from thermal mass is somewhat reduced. We also see from Table~\ref{Tab:tab-1} that reducing $\lambda_\phi$ from its perturbative limit of $\sqrt{4\pi}$ to $1$ reduces the thermal mass effect as well as the NLO effects correspondingly. However, the modification on including the NLO matrix element corrections remain substantial, for example, with  $m_\phi^{\left(0\right)}/m_\chi^{\left(0\right)} \sim 1/3$, we find a $\mathcal{O}(30\%)$ change for both values of $\lambda_\phi$. The final column shows the $\Omega h^2$ difference ratio indicating the thermal NLO matrix element effects, which we see can be as large as $\mathcal{O}(10\%)$ in absolute value.

\section{Summary}
\label{sec:sec5}
To summarize, motivated by earlier studies on thermal NLO corrections to dark matter freeze out in the non-relativistic regime, which found a correction of at most order $(T/m)^2$, we have analysed the NLO thermal corrections in the complementary relativistic regime for DM production from scatterings. In particular, we focussed on an example scenario where DM ($\chi$) is pair produced from the scattering of a relativistic SM-like scalar particle ($\phi$) in the early Universe plasma, through the so-called relativistic freeze-in mechanism. While $\phi$ is taken to be thermalized with the SM sector, the DM remains non-thermal throughout due to its very small couplings. We studied both the vacuum and thermal NLO corrections to the $\phi \phi \rightarrow \chi \chi$ scattering process, where the thermal NLO corrections in this case are driven by $s$-channel $\phi$ particle loops. We also presented approximate analytic formulae for these scattering matrix elements in a near-relativistic regime, and found that the thermal matrix elements typically scale as $T^2/s$ in the relativistic regime for $s$-channel processes.

After exact numerical integration of the NLO scattering rates, we presented a detailed comparison of four different computational setups: using (i) LO matrix elements, (ii) LO matrix elements combined with thermal mass corrections, (iii) NLO vacuum matrix elements coupled with thermal masses, and (iv) full NLO vacuum and thermal matrix elements with in-medium mass corrections. While, (i) and (ii) are often employed in the DM literature for estimating yields, (iv) represents the complete determination of the virtual and thermal NLO effects. We also presented (iii) in order to access the relative contribution of the thermal NLO matrix element corrections with respect to the vacuum NLO effects. In the DM literature, there exist several studies on either vacuum NLO corrections, or thermal NLO corrections separately, but the number of studies combining both NLO effects together are fewer, and our study should be a relevant contribution in this regard.

For the particular freeze-in example considered, we observed that including only the thermal mass effects to LO matrix elements overestimates the reduction in the rate considerably, and the complete NLO correction predicts a scattering rate that falls in between these two approximations. When translated to the DM yield and relic abundance, the corrections due to the NLO matrix elements can be $\mathcal{O}(30\%)$, which is substantial. Within the NLO matrix element corrections, the impact of the thermal NLO correction component on the relic abundance can also be as large as $\mathcal{O}(10\%)$ in absolute value. These conclusions stand for a wide range of values of the scalar self-coupling which enters the NLO corrections, as well as different ratios of the DM and scalar particle masses, as we demonstrate in detail. Therefore, to conclude, vacuum and thermal NLO corrections are highly relevant for accurate determinations of DM abundances produced in relativistic scatterings of bath particles, and should be included for improved theoretical predictions that match the current observational accuracies.

\section*{Acknowledgements}
We would like to thank Deep Ghosh for collaboration at the initial stages of this work, and for many helpful inputs in developing the numerical routines. We also acknowledge Utpal Chattopadhyay for helping with our HPC system. SM is thankful for the hospitality of Mihoko M. Nojiri at KEK, Tsukuba, Japan, where part of this work was carried out, and thanks Kaoru Hagiwara, Kyohei Mukaida and Shigeki Matsumoto for useful discussions. SR thanks Rohan Pramanick for computational help.

\section{Appendix}

We outline here a proof showing how the rate equation can be rearranged in terms of 
matrix elements having only type-1 vertices at the external vertices (vertices connected to 
at-least one external leg). In the following discussion regarding this proof, we interchangeably 
use $+$ and $-$ type to denote type 1 and 2 fields respectively. 

As we are interested in the evolution of the phase space distribution of dark matter (DM) $f_\chi$, 
the most general setup to track this in real time closed time path (CTP) formulation is through 
Kadanoff-Baym equation. In quasi-particle limit after performing Wigner transformation, followed by 
gradient expansion keeping the leading term, it can be shown that the evolution of $f_\chi$ takes 
the form \cite{Beneke:2014gla, Weinstock:2005jw}, 

\begin{align}
\frac{{\rm d}f_\chi(\vec{p})}{{\rm d} t} = \int_0^\infty \frac{{\rm d} p_0}{\pi} \frac{1}{2} 
\left(i \Sigma^>(p)\, iG^<(p) - i \Sigma^<(p)\, iG^>(p) \right) . 
\label{eq:kb1}
\end{align} 
where, in the above expression $\Sigma^>(p)$ and $\Sigma^<(p)$ respectively refer to greater and lesser 
self energy of DM with four momentum $p^\mu = (p^0, \vec{p})$ associated with the external leg. $G^>(p)$ 
and $G^<(p)$ refer respectively to the greater and lesser Green's functions of the theory. As these Green's 
functions can be related to the distribution function and spectral function $\rho(p_0, \vec{p})$, it can be 
shown that the rate equation of $f_\chi$ takes the well-known form \cite{Bellac:2011kqa},

\begin{align}
\frac{{\rm d}f_\chi(\vec{p})}{{\rm d} t} = \int_0^\infty \frac{{\rm d} p_0}{\pi} \frac{\rho(p_0, \vec{p})}{2} 
 \left(- f_\chi(p) \Sigma^>(p) + \left(1+ f_\chi (p) \right)  \Sigma^<(p) \right) . 
\label{eq:kb2}
\end{align} 
As $\Sigma^>(p)$ and $\Sigma^<(p)$ are understood to be related to the annihilation and 
production rates respectively, Eq.~\eqref{eq:kb2} resembles the standard collision 
integral. 

In the model considered in Sec.~\ref{sec:sec2}, $\Sigma^>$ is specified by the sum of the diagrams as shown 
in Eq.~\eqref{Eq:sig_g} upto three loops. The contributions have been grouped as $ \Sigma^{>}_{\rm relevant} $ 
and $ \Sigma^{>}_{\rm remnant} $ for later convenience, where they are given by sum of diagrams respectively 
shown in first and second parentheses. 
\begin{align}
 %  Line 1
 \Sigma^>   
  = & ~ \left( \vcenter{\hbox{\includegraphics[width=0.9\textwidth]{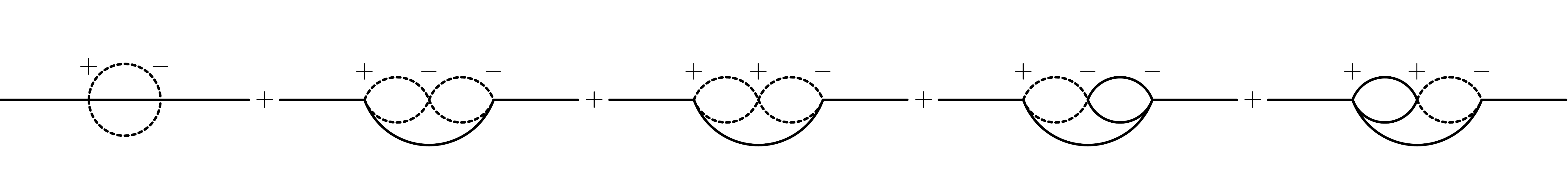}}} \right)  \nonumber \\
  & + \left( \vcenter{\hbox{\includegraphics[width=0.9\textwidth]{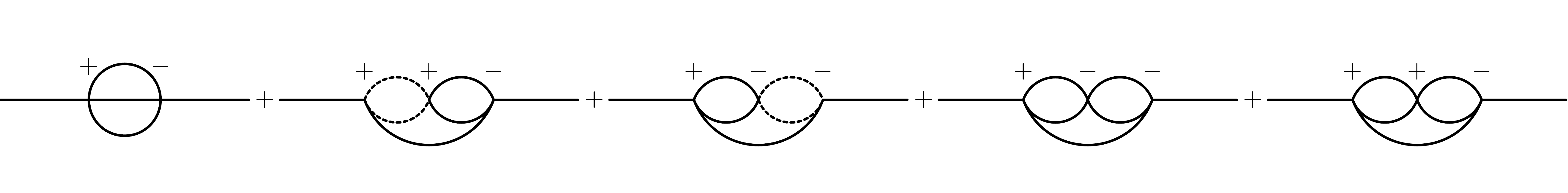}}} \right)  \nonumber \\
  %  Line 3 
  = & ~~  \Sigma^{>}_{\rm relevant} + \Sigma^{>}_{\rm remnant} ~.
\label{Eq:sig_g}
\end{align}

Similarly, diagrams contributing to $\Sigma^<$ till this order are also listed below,
\begin{align}
 %  Line 1
 \Sigma^<   
  = & ~ \left( \vcenter{\hbox{\includegraphics[width=0.9\textwidth]{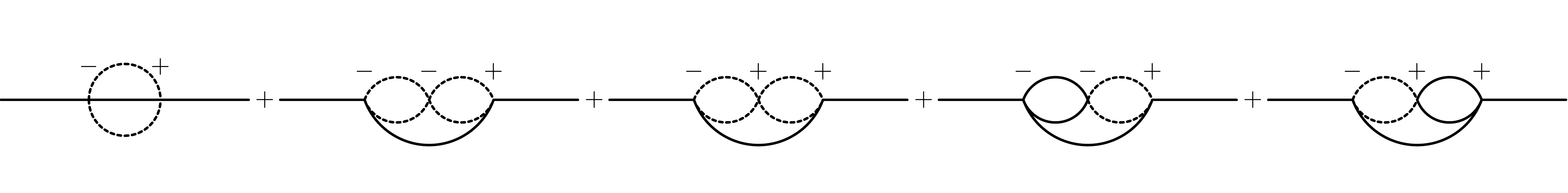}}} \right)  \nonumber \\
  & + \left( \vcenter{\hbox{\includegraphics[width=0.9\textwidth]{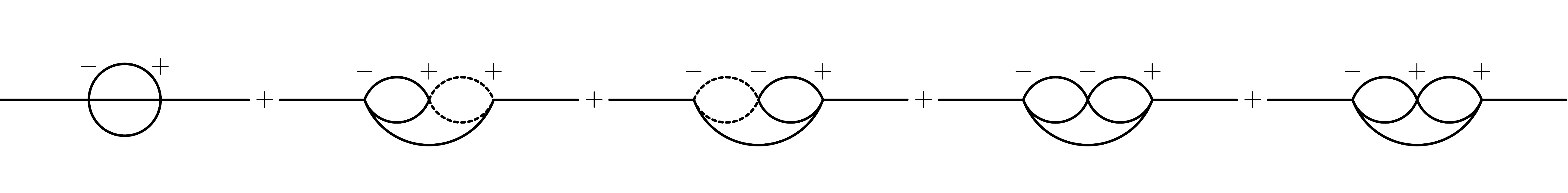}}} \right)  \nonumber \\
  %  Line 3 
  = & ~~  \Sigma^{<}_{\rm relevant} + \Sigma^{<}_{\rm remnant} ~.
\label{Eq:sig_l}
\end{align}

We now proceed with the Feynman diagrammatics to relate the cut diagrams of self energies to corresponding 
matrix elements. After performing proper cuts $\Sigma^>_{\rm relevant}$ is found to be related to DM 
annihilation processes matrix elements as shown below,
\begin{align}
\Sigma^{>}_{\rm relevant} =   & \int {\rm d} \Pi_{\chi \phi \phi} \, f_\chi \left( 1+ f_\phi \right) \left( 1+ f_\phi \right) \left|\mathcal{M}^+_{\rm annihilation}\right|^2   \nonumber \\
  = & \int {\rm d} \Pi_{\chi \phi \phi} \, f_\chi \left( 1+ f_\phi \right) \left( 1+ f_\phi \right)  
  \left|\vcenter{\hbox{\includegraphics[width=5.5cm]{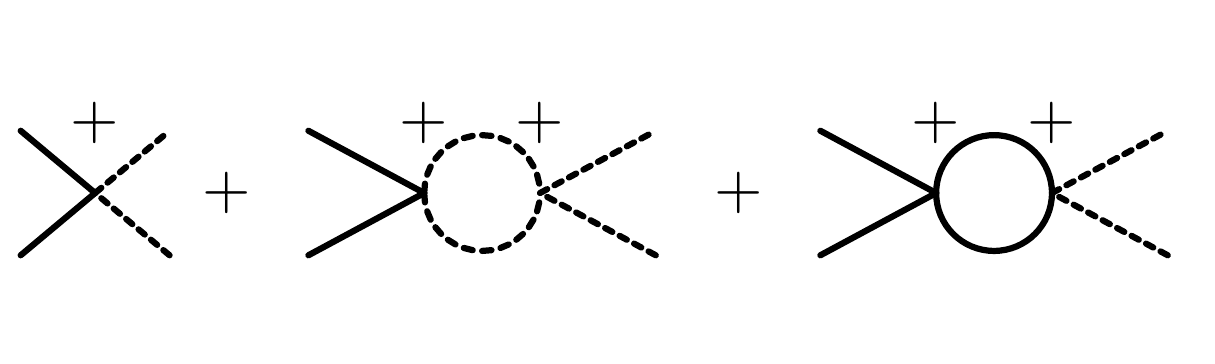}}} \right|^2 . 
\label{Eq:sig_g_rel}
\end{align}

The manipulation in Eq.~\eqref{Eq:sig_g_rel} relies on the fact that, propagators connecting $+$ and $-$ lines 
, \textit{i.e.}, $D_{+-}$ and $D_{-+}$ are on-shell (known in the literature as general on-shell ansatz). 
This then enables one to relate the phase space factors ${\rm d}\Pi_{\rm particles}$, along with 
the distribution functions, exactly to the relevant cut diagrams as shown above. The $+$ sign in the superscript 
of the matrix element $\mathcal{M^+_{\rm annihilation}}$ signifies that all the external vertices of the matrix element is of type-1 (+) only. 

A similar analysis can be performed with 
DM production processes to relate that with $\Sigma^<_{\rm relevant}$, as below,
\begin{align}
\Sigma^{<}_{\rm relevant} = & \int {\rm d} \Pi_{\phi \phi \chi} \, f_\phi  f_\phi \left( 1+ f_\chi \right) 
  \left|\mathcal{M}^+_{\rm production} \right|^2  \nonumber \\
  = & \int {\rm d} \Pi_{\phi \phi \chi} \, f_\phi  f_\phi \left( 1+ f_\chi \right) 
  \left|\vcenter{\hbox{\includegraphics[width=5.5cm]{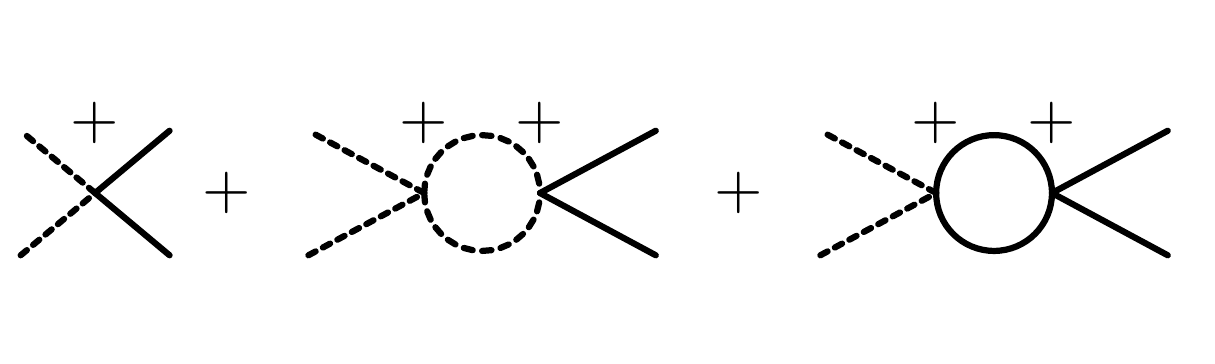}}} \right|^2 .
\label{Eq:sig_l_rel}
\end{align}
It can be shown in a similar way that $\Sigma^{>}_{\rm remnant}$ can be related to 
DM self-scattering matrix element,
\begin{equation}
\Sigma^{>}_{\rm remnant} =  \int {\rm d} \Pi_{\chi \chi \chi} \, f_\chi  \left( 1+ f_\chi \right) \left( 1+ f_\chi \right) \left|\vcenter{\hbox{\includegraphics[width=5.5cm]{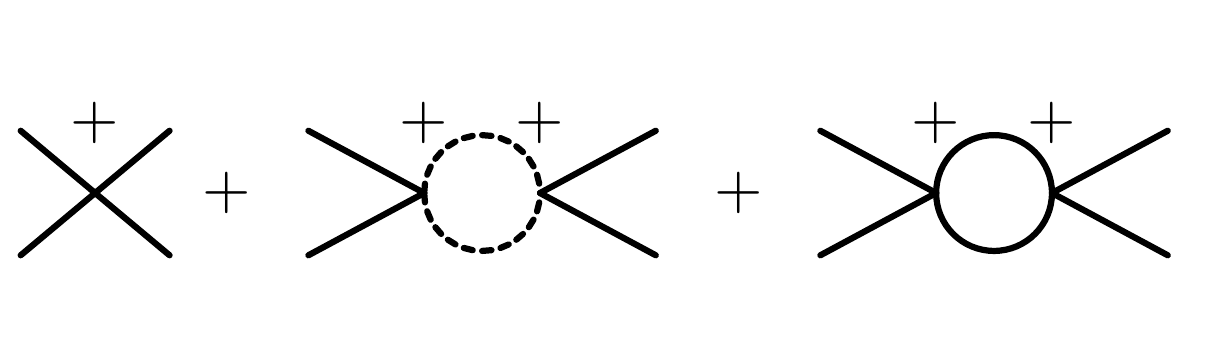}}}\right|^2 .
\label{Eq:sig_g_rem} 
\end{equation}
And, $\Sigma^{<}_{\rm remnant}$ can be expressed as,
\begin{equation}
\Sigma^{<}_{\rm remnant} =  \int {\rm d} \Pi_{\chi \chi \chi} \, f_\chi  f_\chi \left( 1+ f_\chi \right) \left|\vcenter{\hbox{\includegraphics[width=5.5cm]{mat_self}}}\right|^2 .
\label{Eq:sig_l_rem}
\end{equation}
From Eq.~\ref{Eq:sig_g_rem} and Eq.~\ref{Eq:sig_l_rem} it can be shown that the `remnant' parts of 
self energies follow the identity,
\begin{align}
\int \frac{{\rm d}^3 p}{(2\pi)^3} \frac{{\rm d} p_0}{\pi} \frac{\rho(p_0, \vec{p})}{2} \left[ \vphantom{\rule{0pt}{1.5em}} -f_\chi(p) \Sigma^{>}_{\rm remnant}(p) + (1+f_\chi(p)) \Sigma^{<}_{\rm remnant}(p) \right] = 0 .
\label{Eq:remnant_cancel}
\end{align}
This is an important result, as till this order we don't expect the self-interaction of DM to affect the 
number density.

After performing three momentum integration on both sides of Eq.~\ref{eq:kb2}, followed by separating out the self-energies 
in `relevant' and `remnant' pieces and using Eq.~\ref{Eq:sig_g_rel}, Eq.~\ref{Eq:sig_l_rel} 
and  Eq.~\ref{Eq:remnant_cancel} we obtain the rate equation of the number density $n_\chi$ to be, 
\begin{align}
\frac{{\rm d}n_\chi}{{\rm d} t} & = \int \frac{{\rm d}^3 p}{(2\pi)^3} \int_0^\infty \frac{{\rm d} p_0}{\pi} \frac{\rho(p_0, \vec{p})}{2} 
\left( \vphantom{\rule{0pt}{1.5em}} - f_\chi(p) \Sigma^>(p) + \left(1+ f_\chi (p) \right)  \Sigma^<(p) \right) \nonumber \\
& = \int \frac{{\rm d}^3 p}{(2\pi)^3} \frac{1}{2 E_p}  \int {\rm d} \Pi_{\phi \phi \chi}^k \left|\mathcal{M}^+\right|^2 \left[ \vphantom{\rule{0pt}{1.5em}} f_\phi  f_\phi \left( 1+ f_\chi \right)\left(1+ f_\chi (p) \right) \right. \nonumber \\
& \hspace{3cm} \left. \vphantom{\rule{0pt}{1.5em}} -f_\chi(p) f_\chi \left( 1+ f_\phi \right) \left( 1+ f_\phi \right) 
\right] \delta^4\left( p^\chi + k_1^\chi + k_2^\phi + k_3^\phi \right) .
\label{Eq:final_proof}
\end{align}
In the last line of the proof it is assumed that, CP invariance holds, so that we can 
write $\left|\mathcal{M}^+ \right|^2 = \left|\mathcal{M}^+_{\rm production} \right|^2 = \left|\mathcal{M}^+_{\rm annihilation} \right|^2$. 
The spectral function $\rho(p_0, \vec{p})$ is written in quasi-particle, weak coupling limit in the form 
$2 \pi \epsilon(p_0) \delta(p_0^2 - E_p^2)$, where $\epsilon(p_0)$ is the sign function, and $E_p$ satisfies 
the dispersion relation $ E_p^2 = |\vec{p}|^2 + m_T^2 $, where $ m_T $ is the medium modified mass.   
 
We have obtained the well known form of the Boltzmann equation with $\mathcal{M}^+$ denoting the matrix element 
of the concerned process with only type-1 (+) vertices as external vertices. Actually, it has been shown 
in a generic theory that the collision term in the Eqn.~\eqref{eq:kb2} can be rearranged in terms 
of matrix element $\mathcal{M}^+$ squared \cite{Weinstock:2005jw} \footnote{It is noteworthy to mention 
here that, this proof for a generic theory has only been shown to exist for tree level cut diagrams.} alone.

Therefore, in our theory till NLO order, as there can be at-most two external vertices 
(connected to external legs) both of which are of type-1, this only allows for $D_{11}$ propagators. In this 
connection we note that, it has been observed that in the non-relativistic limit involving theories with 
heavy particles, type-2 fields can be shown to effectively decouple \cite{Brambilla:2008cx, Biondini:2013xua}, 
simplifying the calculation. In the scenario of our concern which is a relativistic one, type-2 fields do not 
decouple.  As shown above the intermediate algebra (particularly, complex conjugation along with proper cuts) 
takes care of the contributions of the type-2 fields consistently. Therefore, till the NLO order we are 
interested in, we do not need to explicitly include type-2 fields, the contributions of 
which are consistently accounted through the algebra.

\end{document}